\documentclass[draft,showpacs,showkeys,eqsecnum,nofootinbib,aps]{revtex4}
\renewcommand{\theequation}{\arabic{equation}}
\def\beq{\begin{equation}}
\def\eeq{\end{equation}}
\def\bea{\begin{eqnarray}}
\def\eea{\end{eqnarray}}
\def\nn{\nonumber}

\def\pa{\partial}

\begin{document}
\title{Schr\"odinger representation for topological solitons}
\author{Soon-Tae Hong}
\email{soonhong@ewha.ac.kr}
\affiliation{Department of Science
Education, Ewha Womans University, Seoul 120-750 Korea}
\date{December 31, 2003}%
\begin{abstract}
Exploiting the SU(2) Skyrmion Lagrangian with second-class
constraints associated with Lagrange multiplier and collective
coordinates, we convert the second-class system into the
first-class one in the Batalin-Fradkin-Tyutin embedding through
introduction of the St\"uckelberg coordinates.  In this extended
phase space we construct the ``canonical" quantum operator
commutators of the collective coordinates and their conjugate
momenta to describe the Schr\"odinger representation of the SU(2)
Skyrmion, so that we can define isospin operators and their
Casimir quantum operator and the corresponding eigenvalue equation
possessing integer quantum numbers, and we can also assign via the
homotopy class $\pi_{4}(SU(2))=Z_{2}$ half integers to the isospin
quantum number for the solitons in baryon phenomenology. Different
from the semiclassical quantization previously performed, we
exploit the ``canonical" quantization scheme in the enlarged phase
space by introducing the St\"uckelberg coordinates, to evaluate
the baryon mass spectrum having global mass shift originated from
geometrical corrections due to the $S^{3}$ compact manifold
involved in the topological Skyrmion.  Including ghosts and
anti-ghosts, we also construct Becci-Rouet-Stora-Tyutin invariant
effective Lagrangian.
\end{abstract}
\maketitle

\section{Introduction}

It is well known that baryons can be obtained from topological solutions,
known as SU(2) Skyrmions, since homotopy group $\Pi_{3}(SU(2))=Z$ admits
fermions~\cite{adkins83,sk,hsk}. Using collective coordinates of isospin
rotation of the Skyrmion, Witten and coworkers~\cite{adkins83} have performed
semiclassical quantization having static properties of baryons within 30\%
of the corresponding experimental data. The hyperfine splittings for the
SU(3) Skyrmion~\cite{su3} has been also studied in the SU(3) cranking method
by exploiting rigid rotation of the Skyrmion in the collective
space of SU(3) Euler angles with full diagonalization of the flavor symmetry
breaking terms~\cite{fsbsu3}. Callan and Klebanov~\cite{callan} later suggested
an interpretation of baryons containing a heavy quark as bound states of
solitons of the pion chiral Lagrangian with mesons. In their formalism,
the fluctuations in the strangeness direction are treated differently from
those in the isospin directions.  Moreover, exploiting the standard flavor symmetric
SU(3) Skyrmion rigid rotator approach~\cite{kleb94}, the SU(3) Skyrmion
with the pion mass and flavor symmetry breaking (FSB) terms has been studied to
investigate the chiral symmetry breaking pion mass and FSB effects on the
ratio of the strange-light to light-light interaction strengths and that
of the strange-strange to light-light~\cite{hongprd99}.  However, due to
the geometrical constraints involved in the SU(2) group manifold of the
Skyrmion, the pairs of the collective coordinates and their momenta were
not yet canonical conjugate ones on quantum level and thus these
quantizations could be only semiclassically performed.

On the other hand, the Dirac method~\cite{di} is a well known formalism to
quantize physical systems with constraints. In this method, the Poisson
brackets in a second-class constraint system are converted into Dirac
brackets to attain self-consistency. The Dirac brackets, however, are
generically field-dependent, nonlocal and contain problems related to
ordering of field operators.  To overcome these problems, Batalin,
Fradkin and Tyutin (BFT)~\cite{BFT} developed a method which converts the
second-class constraints into first-class ones by introducing
St\"uckelberg fields.  This BFT embedding has been successively applied to several
models of current interest~\cite{hong02pr}.  Recently, to show novel
phenomenological aspects~\cite{cs}, the compact form of
the first-class Hamiltonian has been constructed~\cite{hong99o3}
for the O(3) nonlinear sigma model, which has been also studied to investigate
the Lagrangian, symplectic, Hamilton-Jacobi and BFT embedding
structures~\cite{rothe03}.  The Becci-Rouet-Stora-Tyutin (BRST)
symmetries~\cite{brst} have been also constructed for constrained
systems~\cite{hong02pr} in the Batalin-Fradkin-Vilkovisky (BFV) scheme~\cite{bfv}.

The motivation of this paper is to quantize ``canonically" the SU(2)
Skyrmion model in the BFT embedding by constructing quantum operator
commutators of the canonical coordinates and their conjugate momenta.
In the Schr\"odinger representation of this system, we will study
the quantum mechanical characteristics to yield the energy
spectrum of the topological Skyrmion, with the geometrical global shift
originated from the compactness of the target manifold.  In section 2, we will
construct the first-class Hamiltonian of the SU(2) Skyrmion in the BFT embedding by
including the St\"uckelberg coordinates.  In section 3, introducing
the quantum commutators in the Schr\"odinger representation
of the SU(2) Skyrmion, we will obtain the baryon mass spectrum having the geometrical
global mass shifts.  In section 4, we will derive the effective Lagrangian
invariant under the BRST symmetries.

\section{First-class Hamiltonian of SU(2) Skyrmion}
\setcounter{equation}{0}
\renewcommand{\theequation}{\arabic{section}.\arabic{equation}}

Now we start with the SU(2) Skyrmion Lagrangian of the form
\begin{equation}
L_{0}=\int{\rm d}r^{3}\left[-\frac{f_{\pi}^{2}}{4}{\rm
tr}(l_{\mu}l^{\mu}) +\frac{1} {32e^{2}}{\rm
tr}[l_{\mu},l_{\nu}]^{2}\right], \label{sklagrangian}
\end{equation}
where $l_{\mu}=U^{\dagger}\partial_{\mu}U$ and $U$ is an SU(2)
matrix satisfying the boundary condition $\lim_{r \rightarrow
\infty} U=I$ so that the pion field vanishes as $r$ goes to
infinity.  In the Skyrmion model, since the hedgehog ansatz has
maximal or spherical symmetry, it is easily seen that spin plus
isospin equals zero, so that isospin transformations and spatial
rotations are related to each other and spin and isospin states
can be treated by collective coordinates $a_{\mu}=(a_{0},\vec{a})$
$(\mu=0,1,2,3)$ corresponding to the spin and isospin rotations
\begin{equation}
A(t) = a_{0}+i\vec{a}\cdot\vec{\tau},
\end{equation}
which is the time dependent collective variable defined on the
SU(2)$_{F}$ group manifold and is related with the zero modes
associated with the collective rotation.  Here $\tau_{i}$
$(i=1,2,3)$ are the Pauli matrices.  With the hedgehog ansatz and
the collective rotation $A(t)\in$ SU(2), the chiral field can be
given by $U(\vec{x},t)=A(t)U_{0}(\vec{x})
A^{\dagger}(t)=e^{i\tau_{a}R_{ab}\hat{x}_{b}f(r)}$ where
$R_{ab}=\frac{1}{2} {\rm tr} (\tau_{a}A\tau_{b}A^{\dagger})$ and
the Skyrmion Lagrangian can be written as
\begin{equation}
L_{0}=-M_{0}+2{\cal I}\dot{a}_{\mu}\dot{a}_{\mu}+a_{4}a_{\mu}\dot{a}_{\mu},
\label{lag}
\end{equation}
where $a_{4}$ is the Lagrange multiplier implementing the
second-class constraint $a_{\mu}\dot{a}_{\mu}\approx 0$ associated
with the geometrical constraint $a_{\mu}a_{\mu}-1\approx 0$ and
$M_{0}$ and ${\cal I}$ are the static mass and the moment of
inertia given as \bea M_{0}&=&\frac{2\pi
f_{\pi}}{e}\int_{0}^{\infty}{\rm d}z~z^{2}\left[\left(
\frac{d\theta}{dz}\right)^{2}+\left(2+2\left(\frac{d\theta}{dz}\right)^{2}
+\frac{\sin^{2}\theta}{z^{2}}\right)\frac{\sin^{2}\theta}{z^{2}}\right],
\label{skstaticmass}\\
{\cal I}&=&\frac{8\pi}{3e^{3}f_{\pi}}\int_{0}^{\infty}{\rm
d}z~z^{2} \sin^{2}\theta
\left[1+\left(\frac{d\theta}{dz}\right)^{2}
+\frac{\sin^{2}\theta}{z^{2}}\right], \label{skinertiamom} \eea
with the dimensionless quantity $z=ef_{\pi}r$.

From the Lagrangian (\ref{lag}) the canonical momenta conjugate to
the collective coordinates $a_{\mu}$ and the Lagrange multiplier $a_{4}$ are given by
\bea
\pi_{\mu}&=&4{\cal I}\dot{a}_{\mu}+a_{\mu}a_{4},\nonumber\\
\pi_{4}&=&0. \label{momenta} \eea
Exploiting the canonical momenta (\ref{momenta}), we then obtain
the canonical Hamiltonian
\begin{equation}
H=M_{0}+\frac{1}{8{\cal I}}(\pi_{\mu}-a_{\mu}a_{4})
(\pi_{\mu}-a_{\mu}a_{4}). \label{canH}
\end{equation}
The usual Dirac algorithm is readily shown to lead to the pair of
second-class constraints $\Omega_{i}$ $(i=1,2)$ as follows \bea
\Omega_{1}&=& \pi_{4}\approx 0,\nonumber\\
\Omega_{2}&=& a_{\mu}\pi_{\mu}-a_{\mu}a_{\mu}a_{4} \approx 0.
\label{const22} \eea to yield the corresponding constraint algebra
with $\epsilon^{12}=-\epsilon^{21}=1$
\begin{equation}
\Delta_{kk^{\prime}}=\{\Omega_{k},\Omega_{k^{\prime}}\}
=\epsilon^{kk^{\prime}}a_{\mu}a_{\mu}. \label{delta}
\end{equation}

Following the BFT embedding~\cite{BFT}, we systematically
convert the second-class constraints $\Omega_i=0$ $(i=1,2)$ into
first-class ones by introducing two St\"uckelberg coordinates
$(\theta, \pi_{\theta})$ with Poisson bracket
\begin{equation}
\{\theta, \pi_{\theta}\}=1. \label{phii}
\end{equation}
The strongly involutive first-class constraints
$\tilde{\Omega}_{i}$ are then constructed as a power series of the
St\"uckelberg coordinates,
\begin{eqnarray}
\tilde{\Omega}_{1}&=&\Omega_{1}+\theta,  \nonumber \\
\tilde{\Omega}_{2}&=&\Omega_{2}-a_{\mu}a_{\mu}\pi_{\theta}.
\label{1stconst}
\end{eqnarray}
Note that the first-class constraints (\ref{1stconst}) can be
rewritten as
\begin{eqnarray}
\tilde{\Omega}_{1}&=&\tilde{\pi}_{4},  \nonumber \\
\tilde{\Omega}_{2}&=&\tilde{a}_{\mu}\tilde{\pi}_{\mu}
-\tilde{a}_{\mu}\tilde{a}_{\mu}\tilde{a}_{4},
\label{oott}
\end{eqnarray}
which are form-invariant with respect to the second-class
constraints (\ref {const22}).

We next construct the first-class variables $\tilde{{\cal F}}
=(\tilde{a}_{\mu},\tilde{\pi}_{\mu})$, corresponding to the
original coordinates defined by ${\cal F}=(a_{\mu},\pi_{\mu})$ in the
extended phase space. They are obtained as a power series in the
St\"uckelberg coordinates $(\theta,\pi_{\theta})$ by demanding that they be
in strong involution with the first-class constraints
(\ref{1stconst}), that is $\{\tilde{\Omega}_{i}, \tilde{{\cal
F}}\}=0$.  After some tedious algebra, we obtain for the
first-class coordinates and their momenta
\begin{eqnarray}
\tilde{a}_{\mu}&=&a_{\mu}\left(\frac{a_{\sigma}a_{\sigma}+2\theta}
{a_{\sigma}a_{\sigma}}\right)^{1/2}
\nonumber \\
\tilde{\pi}_{\mu}&=&\left(\pi_{\mu}+2a_{\mu}a_{4}\frac{\theta}
{a_{\sigma}a_{\sigma}}+2a_{\mu}\pi_{\theta}\frac{\theta}
{a_{\sigma}a_{\sigma}}\right)\left(\frac{a_{\sigma}a_{\sigma}}
{a_{\sigma}a_{\sigma}+2\theta}\right)^{1/2},\nonumber \\
\tilde{a}_{4}&=&a_{4}+\pi_{\theta},\nonumber \\
\tilde{\pi}_{4}&=&\pi_{4}+\theta,
\label{pitilde}
\end{eqnarray}
and the first-class Hamiltonian
\begin{equation}
\tilde{H}=M_{0} +\frac{1}{8{\cal
I}}(\tilde{\pi}_{\mu}-\tilde{a}_{\mu}\tilde{a}_{4})
(\tilde{\pi}_{\mu}-\tilde{a}_{\mu}\tilde{a}_{4}). \label{htilde}
\end{equation}

\section{Schr\"odinger representation for SU(2) Skyrmion}
\setcounter{equation}{0}
\renewcommand{\theequation}{\arabic{section}.\arabic{equation}}

In this section, we start with noting that the first-class coordinates and their momenta
(\ref{pitilde}) are found to satisfy the Poisson algebra
\begin{eqnarray}
\{\tilde{a}_{\mu},\tilde{a}_{\nu}\}&=&0,  \nonumber \\
\{\tilde{a}_{\mu},\tilde{\pi}_{\nu}\}&=&\delta_{\mu\nu},  \nonumber \\
\{\tilde{\pi}_{\mu},\tilde{\pi}_{\nu}\}&=&0,  \label{commst}
\end{eqnarray}
which, in the extended phase space, yield the quantum commutators
as in the cases of unconstrained systems
\begin{eqnarray}
\left[\hat{a}_{\mu},\hat{a}_{\nu}\right]&=&0,  \nonumber \\
\left[\hat{a}_{\mu},\hat{\pi}_{\nu}\right]&=&i\hbar\delta_{ab}, \nonumber \\
\left[\hat{\pi}_{\mu},\hat{\pi}_{\nu}\right]&=&0. \label{commst3}
\end{eqnarray}
We emphasize here that the quantum commutators in (\ref{commst3}) enable
\footnote{In the case of the SU(2) Skyrmion Lagrangian without explicit inclusion
of the constraint $a_{4}a_{\mu}\dot{a}_{\mu}$, we can find the unusual
Poisson algebra $\{\tilde{a}_{\mu},\tilde{\pi}_{\nu}\}=\delta_{\mu\nu}
-\tilde{a}_{\mu}\tilde{a}_{\nu}$ for instance~\cite{hongmpla,hong02pr}, so that strictly speaking
we cannot construct the quantum operator for $\hat{\pi}_{\mu}$ as in (\ref{op-pi}),
since $(\tilde{a}_{\mu},\tilde{\pi}_{\nu})$ are not ``canonical" conjugate pair any more.}
us to describe the Schr\"odinger representation of the SU(2) Skyrmion and to
construct the quantum operator for $\hat{\pi}_{\mu}$
\beq
\hat{\pi}_{\mu}=-i\hbar\frac{\pa}{\pa a_{\mu}}. \label{op-pi}
\eeq
The spin operators $\hat{S}_{i}$ and the isopspin operators $\hat{I}_{i}$ are then given
by~\cite{adkins83}
\bea
\hat{S}_{i}&=&-\frac{i\hbar}{2}\left(\hat{a}_{0}\frac{\pa}{\pa a_{i}}-\hat{a}_{i}\frac{\pa}{\pa a_{0}}
-\epsilon_{ijk}\hat{a}_{j}\frac{\pa}{\pa a_{k}}\right),\nn\\
\hat{I}_{i}&=&-\frac{i\hbar}{2}\left(\hat{a}_{0}\frac{\pa}{\pa
a_{i}}-\hat{a}_{i}\frac{\pa}{\pa a_{0}}
+\epsilon_{ijk}\hat{a}_{j}\frac{\pa}{\pa a_{k}}\right),
\label{spinop} \eea to yield the Casimir operator \beq
\vec{S}^{2}=\vec{I}^{2}=\frac{\hbar^{2}}{4}\left(-\frac{\pa^{2}}{\pa
a_{\mu}\pa a_{\mu}} +3\hat{a}_{\mu}\frac{\pa}{\pa
a_{\mu}}+\hat{a}_{\mu}\hat{a}_{\nu}\frac{\pa}{\pa a_{\mu}}
\frac{\pa}{\pa a_{\nu}}\right). \label{casimir} \eeq Note that the
Casimir operator (\ref{casimir}) is associated with the
three-sphere Laplacian whose eigenvalue equation is of the form
with quantum numbers $l$ (= integers)~\cite{vil68}, \beq
\left(-\frac{\pa^{2}}{\pa a_{\mu}\pa a_{\mu}}
+3\hat{a}_{\mu}\frac{\pa}{\pa
a_{\mu}}+\hat{a}_{\mu}\hat{a}_{\nu}\frac{\pa}{\pa a_{\mu}}
\frac{\pa}{\pa a_{\nu}}\right)\psi_{l}(A) =l(l+2)\psi_{l}(A). \eeq
We can then find the eigenvalue for the Casimir operator
(\ref{casimir}) as follows \beq
\hat{I}^{2}\psi_{I}(A)=\hbar^{2}I(I+1)\psi_{I}(A), \label{jm} \eeq
where $I=l/2$ are the total isospin quantum numbers of baryons.
Note that not all of the states discussed above are physically
allowed.  Since our solitons are to be fermions, we have the
condition on the quantum wave function~\cite{fin},
\beq
\psi_{I}(-A)=-\psi_{I}(A), \label{psi} \eeq whose allowed values
of $I$ are $I=1/2,~3/2,\cdots$, corresponding to nucleons
$(I=S=1/2)$ and deltas $(I=S=3/2)$ in the baryon
phenomenology~\cite{adkins83,hong02pr}.  Note that the condition
(\ref{psi}) is closely related to the nontrivial homotopy class
$\pi_{4}(SU(2))=Z_{2}$ associated with a space-time manifold
compactified to be $S^{4}=S^{3}\times S^{1}$ where $S^{3}$ and
$S^{1}$ are compactified Euclidean three-space and time,
respectively~\cite{su3,hong02pr}.

Next, following symmetrization procedure~\cite{lee81,hong99sk,hong02pr} together
with ({\ref{htilde}) and (\ref{op-pi}), we arrive at the
Hamiltonian quantum operator for the SU(2) Skyrmion
\bea
\hat{H}&=&M_{0} +:\frac{1}{8{\cal I}}\left(-i\hbar\frac{\pa}{\pa a_{\mu}}
+i\hbar\hat{a}_{\mu}\hat{a}_{\nu}\frac{\pa}{\pa a_{\mu}}\right)\left(-i\hbar\frac{\pa}{\pa a_{\mu}}
+i\hbar\hat{a}_{\mu}\hat{a}_{\nu}\frac{\pa}{\pa a_{\mu}}\right):\nonumber\\
&=&M_{0}+\frac{\hbar^{2}}{8{\cal I}}\left(-\frac{\pa^{2}}{\pa a_{\mu}\pa a_{\mu}}
+3\hat{a}_{\mu}\frac{\pa}{\pa a_{\mu}}+\hat{a}_{\mu}\hat{a}_{\nu}\frac{\pa}{\pa a_{\mu}}
\frac{\pa}{\pa a_{\nu}}+\frac{5}{4}\right)\nn\\
&=&M_{0}+\frac{1}{2{\cal I}}\left(\vec{I}^{2}+\frac{5\hbar^{2}}{16}\right).
\label{op-htilde} \eea
Note that the Hamiltonian quantum operator
(\ref{op-htilde}) has terms of orders $\hbar^{0}$ and $\hbar^{2}$
only, so that one can have static mass (of order $\hbar^{0}$) and
rotational energy contributions (of order $\hbar^{2}$) without any
vibrational mode ones (of order $\hbar^{1}$).  In fact, the
starting Lagrangian (\ref{lag}) does not possess any vibrational
degrees of freedom in itself since it has the kinetic term
describing the motions of the soliton residing on the $S^{3}$
manifold.  The Schr\"odinger representation for the SU(2) Skyrmion can be then
given by the following eigenvalue equation
\beq
\left[M_{0}+\frac{1}{2{\cal I}}\left(\vec{I}^{2}+\frac{5\hbar^{2}}{16}\right)
\right]\psi_{I}=E_{I}\psi_{I},
\label{sch}\eeq
to yield the mass spectrum of the baryons
\beq
E_{I}=M_{0}+\frac{\hbar^{2}}{2{\cal
I}}\left[I(I+1)+\frac{5}{16}\right],
\label{spec}
\eeq
which originate from the static mass and rotational energy contributions
discussed above.

Now, it seems appropriate to discuss the global mass shift
involved in the rotational energy contributions to the mass
spectrum (\ref{spec}).  In fact, in Ref.~\cite{adkins83} the mass
spectrum of the SU(2) Skyrmion was constructed in the framework of
the semiclassical quantization where they ignored the effects of
the geometrical constraints involved in the model.  The mass
spectrum of the SU(2) Skyrmion was later improved with the Weyl
ordering corrections~\cite{neto,hong99sk} in the BFT embedding,
where one still could not construct well-defined quantum
commutators among the collective coordinates and their conjugate
momenta due to the unusual Poisson algebra of these variables.  In
that sense the modified Skyrmion mass spectrum in the previous BFT
embedding was not rigorously constructed.  However, in the above
BFT embedding approach to the SU(2) Skyrmion, we recall that the
quantum commutators among the collective coordinates and their
conjugate momenta are canonically well defined and thus the mass
spectrum has been rigorously evaluated to yield the global mass
shift as in (\ref{spec}).  Note that the global mass shift in the
spectrum (\ref{spec}) originates from the geometrical corrections
due to the characteristics of the $S^{3}$ compact manifold
involved in the topological Skyrmion model.

\section{BRST symmetries in SU(2) Skyrmion}
\setcounter{equation}{0}
\renewcommand{\theequation}{\arabic{section}.\arabic{equation}}

In order to investigate the BRST symmetries~\cite{brst} associated
with the Lagrangian (\ref{lag}) of the SU(2) Skyrmion
model,\footnote{The BRST symmetries were constructed in the SU(2)
Skyrmion Lagrangian which is different from (\ref{lag}) in the
sense that the constraints $a_{\mu}\dot{a}_{\mu}$ is not
explicitly included~\cite{hong02pr,hongmpla}.  The BRST
transformation rules for the Lagrange multiplier and its momentum
$(a_{4},\pi_{4})$ are then missing in the
Refs.~\cite{hong02pr,hongmpla}.} we rewrite the first-class
Hamiltonian (\ref{htilde}) in terms of original collective
variables and St\"uckelberg coordinates \beq
\tilde{H}=M_{0}+\frac{1}{8{\cal
I}}(\pi_{\mu}-a_{\mu}a_{4}-a_{\mu}\pi_{\theta})
(\pi_{\mu}-a_{\mu}a_{4}-a_{\mu}\pi_{\theta})
\frac{a_{\nu}a_{\nu}}{a_{\nu}a_{\nu}+2\theta},  \label{hct} \eeq
which is strongly involutive with the first-class constraints,
$\{\tilde{\Omega}_{i},\tilde{H}\}=0$. Note that with this
Hamiltonian (\ref{hct}), we cannot generate the first-class Gauss'
law constraint from the time evolution of the constraint
$\tilde{\Omega}_{1}$.  By introducing an additional term
proportional to the first-class constraints $\tilde{\Omega}_{2}$
into $\tilde{H}$, we obtain an equivalent first-class Hamiltonian
\begin{equation}
\tilde{H}^{\prime}=\tilde{H}+\frac{1}{4{\cal
I}}\pi_{\theta}\tilde{\Omega}_{2}\label{hctp}
\end{equation}
to generate the Gauss' law constraint \beq
\{\tilde{\Omega}_{1},\tilde{H}^{\prime}\}=\frac{1}{4{\cal
I}}\tilde{\Omega}_{2},~~~
\{\tilde{\Omega}_{2},\tilde{H}^{\prime}\}=0. \eeq Note that these
Hamiltonians $\tilde{H}$ and $\tilde{H}^{\prime}$ effectively act
on physical states in the same way since such states are
annihilated by the first-class constraints.

In the framework of the BFV formalism~\cite{bfv}, by  introducing two
canonical sets of ghosts and anti-ghosts, together with
Lagrange multipliers
\beq
({\cal C}^{i},\bar{{\cal P}}_{i}),~~~
({\cal P}^{i},\bar{{\cal C}}_{i}),~~~
(N^{i},B_{i}),~~~(i=1,2),
\eeq
and the unitary gauge choice $\chi^{1}=\Omega_{1},~~~\chi^{2}=\Omega_{2}$,
we now construct the nilpotent BRST charge $Q$, the fermionic gauge fixing
function $\Psi$ and the BRST invariant minimal Hamiltonian $H_{m}$
\begin{eqnarray}
Q&=&{\cal C}^{i}\tilde{\Omega}_{i}+{\cal P}^{i}B_{i},
\nonumber \\
\Psi&=&\bar{{\cal C}}_{i}\chi^{i}+\bar{{\cal P}}%
_{i}N^{i},  \nonumber \\
H_{m}&=&\tilde{H}+\frac{1}{4{\cal
I}}\pi_{\theta}\tilde{\Omega}_{2} -\frac{1}{4{\cal I}}{\cal
C}^{1}\bar{{\cal P}}_{2}, \label{hmham}
\end{eqnarray}
with the properties $Q^{2}=\{Q,Q\}=0$ and $\{\{\Psi,Q\},Q\}=0$.
The nilpotent charge $Q$ is the generator of the following
infinitesimal transformations, \beq
\begin{array}{lll}
\delta_{Q}a_{\mu}=-{\cal C}^{2}a_{\mu},
&~~\delta_{Q}a_{4}=-{\cal C}^{1},
&~~\delta_{Q}\theta={\cal C}^{2}a_{\mu}a_{\mu},\\
\delta_{Q}\pi_{\mu}={\cal C}^{2}(\pi_{\mu}-2a_{\mu}a_{4}-2a_{\mu}\pi_{\theta}),
&~~\delta_{Q}\pi_{4}=-{\cal C}^{2}a_{\mu}a_{\mu},
&~~\delta_{Q}\pi_{\theta}={\cal C}^{1},\\
\delta_{Q}\bar{{\cal C}}_{i}=B_{i}, &~~\delta_{Q}{\cal C}^{i}=0, &~~\delta_{Q}B_{i}=0,\\
\delta_{Q}{\cal P}^{i}=0, &~~\delta_{Q}\bar{{\cal
P}}_{i}=\tilde{\Omega}_{i},
&~~\delta_{Q}N^{i}=-{\cal P}^{i},\\
\end{array}
\label{brstgaugetrfm} \eeq which in turn imply $\{Q,H_{m}\}=0$,
that is,  $H_{m}$ in (\ref{hmham}) is  the BRST invariant.

After some algebra, we arrive at the effective quantum Lagrangian
of the form
\begin{equation}
L_{eff}=L_{0} + L_{WZ} + L_{ghost} \label{lagfinal}
\end{equation}
where $L_{0}$ is given by (\ref{lag}) and
\begin{eqnarray}
L_{WZ}&=& \frac{4{\cal
I}\theta}{1-2\theta}\dot{a}_{\mu}\dot{a}_{\mu}
-\frac{2{\cal I}}{(1-2\theta)^{2}}\dot{\theta}^{2},\nonumber\\
L_{ghost}&=&-2{\cal I}(1-2\theta)^{2}(B+2\bar{{\cal C}}{\cal
C})^{2} -\frac{\dot{\theta}\dot{B}}{1-2\theta} +\dot{\bar{{\cal
C}}}\dot{{\cal C}}-\frac{(1-2\theta)^{2}}{2}a_{4}(B+2\bar{{\cal
C}}{\cal C}), \label{lagwz}
\end{eqnarray}
with redefinition ${\cal C}={\cal C}^{2}$, $\bar{\cal C}=\bar{\cal C}_{2}$ and
$B=B_{2}$.  This Lagrangian is invariant under the BRST transformation \beq
\begin{array}{lll}
\delta_{\epsilon}a_{\mu}=\epsilon a_{\mu}{\cal C},
&~~\delta_{\epsilon}a_{4}=-2\epsilon a_{4}{\cal C},
&~~\delta_{\epsilon}\theta=-\epsilon
a_{\mu}a_{\mu}{\cal C},\\
\delta_{\epsilon}\bar{{\cal C}}=-\epsilon B, &~~\delta_{\epsilon}{\cal C}=0,
&~~\delta_{\epsilon}B=0,\\
\end{array}
\eeq where $\epsilon$ is an infinitesimal Grassmann valued
parameter.

\section{Conclusion}

In conclusion, we have introduced the  SU(2) Skyrmion Lagrangian
having constraints of the form $a_{4}a_{\mu}\dot{a}_{\mu}$ with
the Lagrange multiplier $a_{4}$ and the collective coordinates
$a_{\mu}$ to convert the second-class Hamiltonian into the
first-class one in the BFT embedding. In this embedding by
including the St\"uckelberg coordinates we have enlarged the phase
space, where we could construct the quantum operator commutators
of the ``canonical" collective coordinates and their conjugate
momenta and then we could describe the Schr\"odinger
representation of the SU(2) Skyrmion model.

Constructing the isospin operators in the enlarged phase space, we
have associated their Casimir quantum operator with the
three-sphere Laplacian to obtain the corresponding eigenvalue
equation having the integer quantum numbers. Classifying the
physical states relevant to these quantum numbers via the homotopy
class $\pi_{4}(SU(2))=Z_{2}$, we have assigned half integers to
the isospin quantum number for the solitons, which are to be
fermionic. Note that, different from the previous results in which
the quantization was semiclassically performed, we have exploited
the ``canonical" quantization scheme in the enlarged phase space
by introducing the St\"uckelberg coordinates degrees of freedom to
evaluate the mass spectrum of the baryons, which has the global
mass shift originated from the geometrical corrections due to the
characteristics of the $S^{3}$ compact manifold involved in the
topological Skyrmion.

Enlarging further the phase space by including the ghosts and anti-ghosts, we
have constructed the nilpotent BRST charge in the BFV formalism to derive the effective
Lagrangian invariant under the BRST transformation associated with the BRST charge.
Here one notes that in constructing the BRST symmetries in the topological SU(2)
Skyrmion having the second-class geometrical constraints it was crucial to introduce
the St\"uckelberg coordinates degrees of freedom, which enable us to exploit the
BFV scheme.

\acknowledgments The author would like to acknowledge financial support
in part from the Korea Science and Engineering Foundation Grant R01-2000-00015.

\end{document}